\documentclass[aps,prl,reprint,superscriptaddress]{revtex4-1}

\usepackage[]{graphicx}
\usepackage{epsfig}
\usepackage{color}

\begin{document}


\title{Exploring small energy scales  with x-ray absorption and dichroism   
}



\author{C.~Praetorius}
\author{M.~Zinner}
\affiliation{Physikalisches Institut, Universit{\"{a}}t W{\"{u}}rzburg, Am Hubland, 97074 W{\"{u}}rzburg, Germany}

\author{P. Hansmann}
\affiliation{Max Planck Institute for Solid State Research, Heisenbergstra\ss e 1, 70569 Stuttgart, Germany}

\author{M. W. Haverkort}
\affiliation{Max Planck Institute for Chemical Physics of Solids, N{\"{o}}thnizer Stra\ss e 40, 01187 Dresden, Germany}

\author{K.~Fauth}
\email[]{fauth@physik.uni-wuerzburg.de}
\affiliation{Physikalisches Institut, Universit{\"{a}}t W{\"{u}}rzburg, Am Hubland, 97074 W{\"{u}}rzburg, Germany}
\affiliation{Wilhelm Conrad R{\"{o}}ntgen-Center for Complex Material Systems (RCCM), Universit{\"{a}}t W{\"{u}}rzburg, Am Hubland, 97074 W{\"{u}}rzburg, Germany}


\date{\today}

\begin{abstract}
Soft x-ray linear and circular dichroism (XLD, XMCD) experiments at the Ce M$_{4,5}$ edges are being used to determine the energy scales characterizing the Ce $4f$ degrees of freedom in the ultrathin ordered surface intermetallic CeAg$_x$/Ag(111). 
We find that all relevant interactions, i.~e.~Kondo scattering, crystal field splitting and magnetic exchange coupling occur on small scales.
Our study demonstrates the usefulness of combining x-ray absorption experiments probing linear and circular dichroism owing to their
strong sensitivity for anisotropies in both charge distribution and pa\-ra\-mag\-ne\-tic response, respectively. 
\end{abstract}


\maketitle


Rare earth intermetallic compounds display a rich phenomenology of physical properties, encompassing very different kinds of ground states, 
such as magnetic order, unconventional superconductivity and paramagnetic heavy fermion liquids \cite{Gege08a,Gege15a}.
The interaction of localized $4f$ electrons with itinerant electronic degrees of freedom may result in the emergence of small characteristic energy scales
which produce nontrivial macroscopic behavior at low temperature and complex phase diagrams with competing interactions and orders \cite{grew91a,Lohn07a,Gege08a,Gege15a,yang08a}.
In a solid environment, the degeneracy of the rare earth $4f$ ground configuration is lifted by the crystal field in general, causing both an anisotropic $4f$ charge distribution and, 
in conjunction with spin orbit coupling, (single ion) magnetic anisotropy.
Unraveling the crystal field induced level structure thus constitutes an essential part of understanding the low temperature physics and of
establishing correlations between local $4f$ symmetry at low temperature on the one hand and macroscopic ground state properties on the other.

In this respect, the usefulness of probing the $4f$ configuration with linear polarized soft x-rays \cite{Sacc92a} has been demonstrated for a variety of Ce compounds in recent years
 \cite{Hans08a, Will09a,Will10a,Will11a,Will12a,Will12b,Strig12a,Will15a},
allowing to settle several open issues, where other experiments left room for diverging interpretations.
Its magnetic variants, x-ray magnetic linear and circular dichroism (XMLD, XMCD) constitute sensitive element and orbital specific probes of magnetic polarization and anisotropy \cite{Schu87a,Laan86b,Chen95a,stoe99a,vanE97a,vand08a}.
XMCD was successfully utilized to reveal the presence of magnetic Kondo screening in CePt$_5$/Pt(111) \cite{Prae15a}.

In the present letter we demonstrate that the combined use of linear and magnetic circular dichroism allows us to determine the crystal field structure 
without recourse to e.~g.~inelastic neutron scattering, as in some previous work \cite{Hans08a,Will09a,Will10a}.
Our chosen example of an ultrathin Ce-Ag surface intermetallic furthermore highlights a threefold advantage of this approach.
First, linear and circular dichroism experiments are frequently both feasible with the same installation
and therefore can be performed \textit{in situ} within a single experimental run.
Second, the splittings turn out to be of the order of $1$ meV only, making their discrimination from
quasi-elastic scattering a difficult task.
Last but not least, the sample volume is so small that most alternative methods would face serious sensitivity challenges.

The formation of an ordered intermetallic phase upon depositing minute amounts of Ce onto Ag(111) held at elevated temperature has been reported before \cite{Schw12a}.
In the preparation of our specimens we have adopted a similar procedure.
In brief, clean Ag(111) was prepared by cycles of Ar$^+$ ion sputtering ($E_{\text{kin}}$: 1 keV) and subsequent annealing to $920$ K.
The crystal was then held at $840$ K while a Ce dose of 
approx.~$1 \times10^{15}$ atoms/cm$^2$ 
was deposited onto it from a thoroughly outgassed W crucible mounted in a commercial electron beam evaporator.


\begin{figure} 
\includegraphics[width=8.5cm]{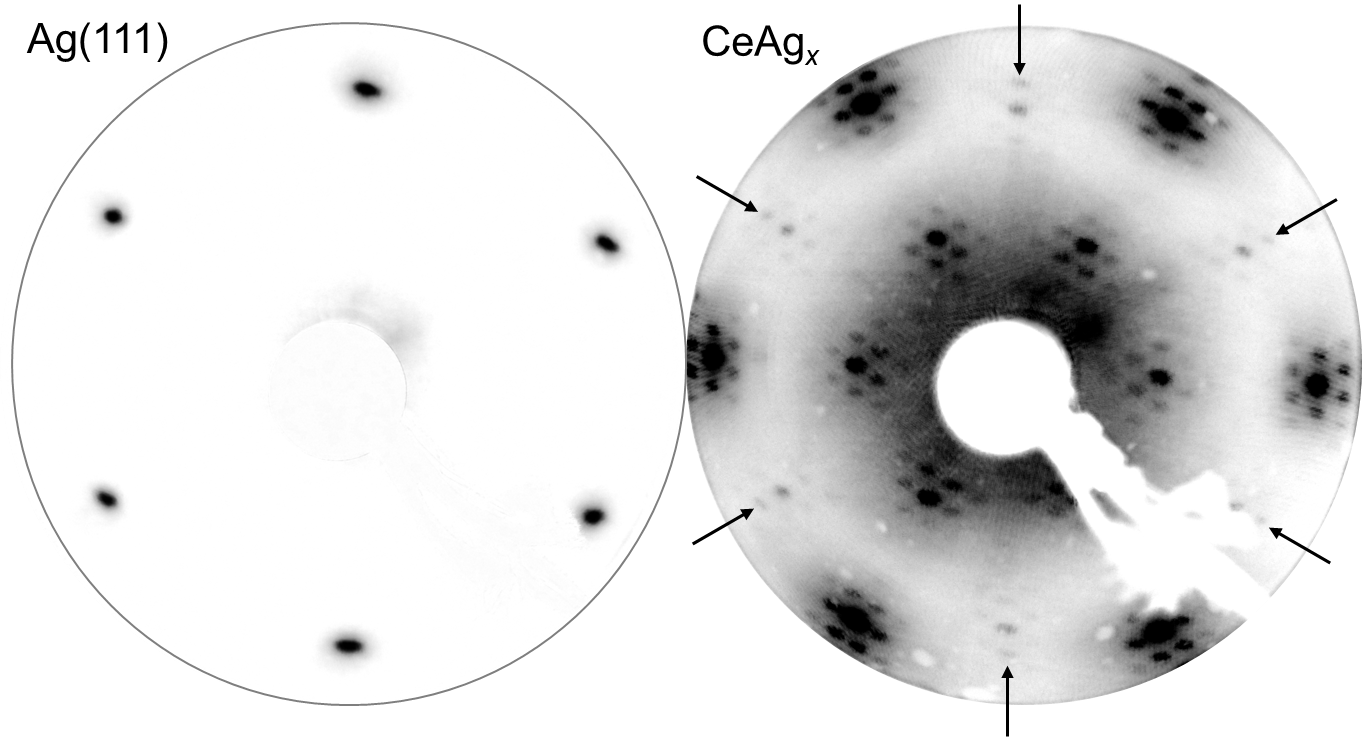} 
\caption{\label{fig1LEED}
LEED patterns recorded at an electron kinetic energy of $70$ eV from Ag(111) (left) and CeAg$_x$ (right).
The satellite reflections in the latter indicate the formation of a $(2\sqrt{3} \times 2\sqrt{3})R30^\circ$ superstructure as in CePt$_5$/Pt(111) at comparable Ce dose \cite{Kemm14a,Prae15a}.
Ag(111) reflections merge with one of the satellite reflections around $\langle 2 \; \overline{1} \rangle_{\text{CeAg}_x}$ (indicated by arrows), yielding a $(\frac{10}{9} \sqrt{3} \times \frac{10}{9}\sqrt{3})R30^\circ$
relation between CeAg$_x$ and substrate unit cells, respectively. 
}
\end{figure}

Fig.~\ref{fig1LEED} displays a LEED pattern of a CeAg$_x$ specimen characteristic of this range of Ce 
coverage next to the one of pristine Ag(111), taken at the same electron kinetic energy.
The diffraction pattern is very much reminiscent of our earlier observations for CePt$_5$/Pt(111) at similar Ce dose \cite{Kemm14a,Prae15a}.
It reveals a combination of two superstructures on two different length scales.
The main diffraction features may be attributed to an intermetallic $(1.1\sqrt{3} \times 1.1\sqrt{3})R30^\circ$ surface reconstruction, whereas the satellites
indicate the formation of a longer range surface corrugation of $(3\sqrt{3} \times 3\sqrt{3})R30^\circ$ character with respect to this intermetallic phase.
This corresponds to a hexagonal surface corrugation in rotational alignment with the substrate lattice and a periodicity of approx.~$15$ nm.
As indicated by arrows in Fig.~\ref{fig1LEED} the Ag(111) first order diffraction beams superimpose with the outermost satellite reflections surrounding the $\langle 2\, \overline{1} \rangle$ spots of the surface intermetallic. 
Like in CePt$_5$ the satellite intensities strongly loose intensity as the initial Ce coverage is increased \cite{Schw12a,Zinnerunpub}.
Despite the obvious similarities between both systems, there are also some differences.
We have so far been unable to determine
the exact composition and structure of this ordered Ce-Ag phase which we therefore label as CeAg$_x$.
A more extensive account of the properties of CeAg$_x$/Ag(111) and their dependence on Ce coverage shall be given in a separate publication \cite{Zinnerunpub}.

For the purpose of this letter it is sufficient to recognize the formation of a hexagonal structure and we shall therefore analyze our results
by assuming sixfold rotational symmetry about the Ce sites.
It is a fundamental property of hexagonal crystal fields (CF) to split the atomic Ce $4f^1$ configuration ($j=5/2$) into three Kramers doublets of pure $m_j$ character.
Unlike in cubic or tetragonal symmetry \cite{Hans08a,Will12a,Will12b} the CF is therefore fully specified by $|m_j\rangle$ level splittings and ordering.
We denote the CF splittings as $\Delta_1 = E_{3/2} - E_{1/2}$ and $\Delta_2 = E_{5/2} - E_{1/2}$ and determine their magnitude and sign from
linear and magnetic circular dichroism measurements at the soft x-ray Ce M$_{4,5}$ edges in what follows.

\begin{figure} 
\includegraphics[width=8.5cm]{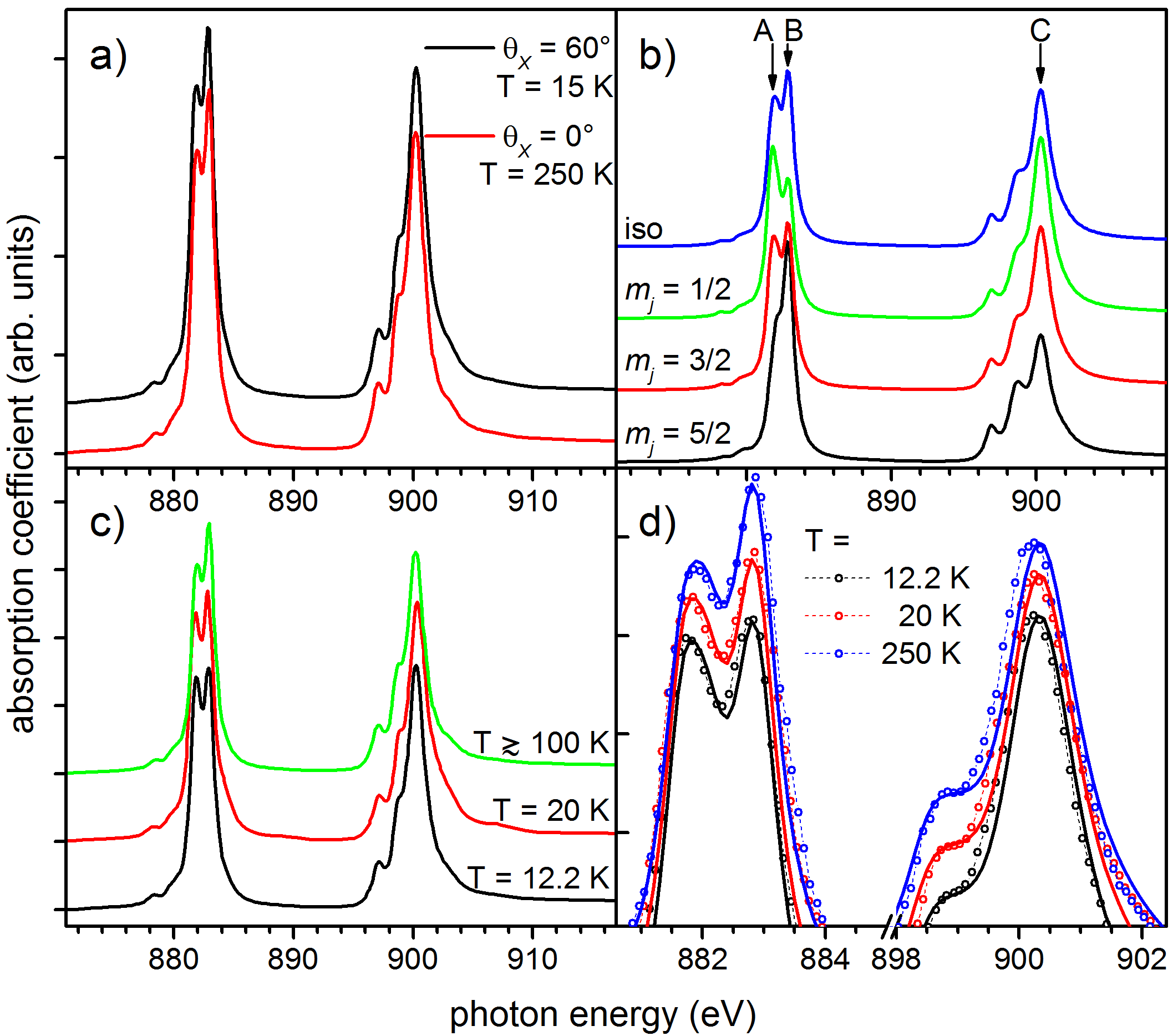} 
\caption{\label{fig2XAS}
Selection of experimental and simulated Ce M$_{4,5}$ XA spectra for CeAg$_x$/Ag(111).
a) Experimental low temperature ($T=15$ K) isotropic and high temperature ($T=250$ K) normal incidence spectrum
b) Simulated isotropic spectrum and normal incidence spectra for pure $|m_j\rangle$ initial states
c) Experimental normal incidence spectra at various temperatures
d) Zoom-in on the main spectral features of both experimental and simulated normal incidence spectra for three temperatures (see text for details).
In each panel, individual spectra were displaced along the ordinate for clarity.
}
\end{figure}

All soft x-ray absorption experiments were carried out at the PM3 bending magnet beamline of BESSY-II, Berlin, using circular polarization ($p\approx 0.93$) \cite{Kach15a}.
X-ray absorption was measured in the total electron yield  (TEY) mode and normalized by the TEY captured from a gold mesh.
Although with reduced amplitude, linear dichroism can nevertheless be observed by variation of the x-ray angle of incidence $\theta_X$.
The polarization averaged, so-called isotropic spectrum is well approximated by oblique incidence data taken at $\theta_X = 60^\circ$ in the present work.
Its line shape is independent of the thermal occupation of the CF states, since it is identical for all $| m_j \rangle$ initial states.
In contrast, spectra measured at normal incidence (NI, i.e. along the hexagonal symmetry axis) do exhibit temperature dependent line shapes,
determined by the fractional occupation of the CF split $|m_j\rangle$ states.
Introducing the Boltzmann weights $p_{1,2} = \exp(-\Delta_{1,2}/k_BT)$, the NI spectrum $I^{NI}(T)$ is given by
\begin{equation} \label{NIspecT}
I^{NI}(T) = Z^{-1}\left ( I^{NI}_{|1/2\rangle} + p_1 I^{NI}_{|3/2\rangle} + p_2 I^{NI}_{|5/2\rangle} \right )   , 
\end{equation}
where $Z=1 + p_1 + p_2$ is the partition function.
Evidently, in the limit $k_BT \gg \Delta_1,\Delta_2$ the line shape observed at NI converges to the isotropic one.
The two experimental spectra displayed in Fig.~\ref{fig2XAS}a) demonstrate that this condition is fulfilled in our CeAg$_x$/Ag(111)
specimens.
The NI spectrum ($\theta_X=0^\circ$) measured at $T=250$ K is hardly distinguishable from the isotropic spectrum.


In our analysis we make use of simulated absorption spectra to trace the experimental temperature dependences.
These simulations were obtained from full atomic multiplet calculations as implemented in the Quanty Package \cite{Have12a,quantylink}.
To obtain the best match between calculated and experimental isotropic spectra, the $ff$ $(df)$ Slater Integrals were reduced by 42.5\% (17.5\%) from their respective Hartree-Fock values,
well in accordance with previous work \cite{Hans08a,Will10a,Will12a,Stri13a,Will15a}.
In addition, the Ce $3d$ core hole spin orbit coupling constant $\zeta_{3d}$ was slightly readjusted to reproduce the experimental separation between the M$_4$ and M$_5$ edges.
Theoretical line spectra were convoluted with a Gaussian (FWHM $0.2$ eV) representing the experimental energy resolution
as well as with Lorentzian contributions to account for the lifetime of the core excited states.
Since the spectral shape of the M$_4$ edge is affected by autoionization decay \cite{Thol85a}, its lifetime broadening was calculated by convolution with a Fano profile ($q\approx 16$) \cite{Schi11a} rather than a Lorentzian.

The resulting Ce M$_{4,5}$ absorption spectra are displayed in Fig.~\ref{fig2XAS}b).
While the overall agreement of the isotropic spectrum with measured data is very good we note that not all multiplet terms can simultaneously be made to coincide with the experimental features 
when applying universal scaling factors to the Slater Integrals.
This is most apparent for the weak shoulder at $903$ eV, which is not discernible in the calculated spectra since its separation from the main M$_4$ peak (feature C) is too small.

The remaining spectra in Fig.~\ref{fig2XAS}b) demonstrate the different spectral shapes owing to the anisotropic charge distribution in the $|m_j\rangle$ states.
In particular, their most prominent peaks (A, B \& C) feature considerable variations in their relative intensities.

Experimental NI data acquired at $T \gtrsim 100$ K display variations which cannot be resolved on the scale of Fig.~\ref{fig2XAS}c).
In comparison to the isotropic spectra, they nevertheless systematically exhibit a slightly larger ($\approx 1$-$2\%$) C/B peak intensity ratio.
Referring to the $|m_j\rangle$ specific spectra of Fig.~\ref{fig2XAS}b) this observation immediately reveals that $|5/2\rangle$ must be an excited state.
As the temperature is lowered, the C/B peak intensity ratio is further enhanced, but in addition the M$_5$ line shape now acquires a noticeable change in spectral appearance.
The observed spectral variations restrict the parameter $\Delta_2$ to a relatively narrow energy window of $\Delta_2= 1.1\pm0.2$ meV.
The determination of the other CF parameter ($\Delta_1$) on the basis of NI XAS data alone is less obvious.
Scenarios with $- 1$ meV $\lesssim \Delta_2 \lesssim 5$ meV can be made to satisfactorily match the sequence of experimental spectra.
This is largely owed to the smallness of the linear dichroism associated with the $|3/2\rangle$ fraction of the initial state.
The choice of parameters $\Delta_1$ and $\Delta_2$ for the simulations in Fig.~\ref{fig2XAS}d) therefore already accounts for the information gained from considering the XMCD signal
which we shall discuss next.


\begin{figure} 
\includegraphics[width=8.5cm]{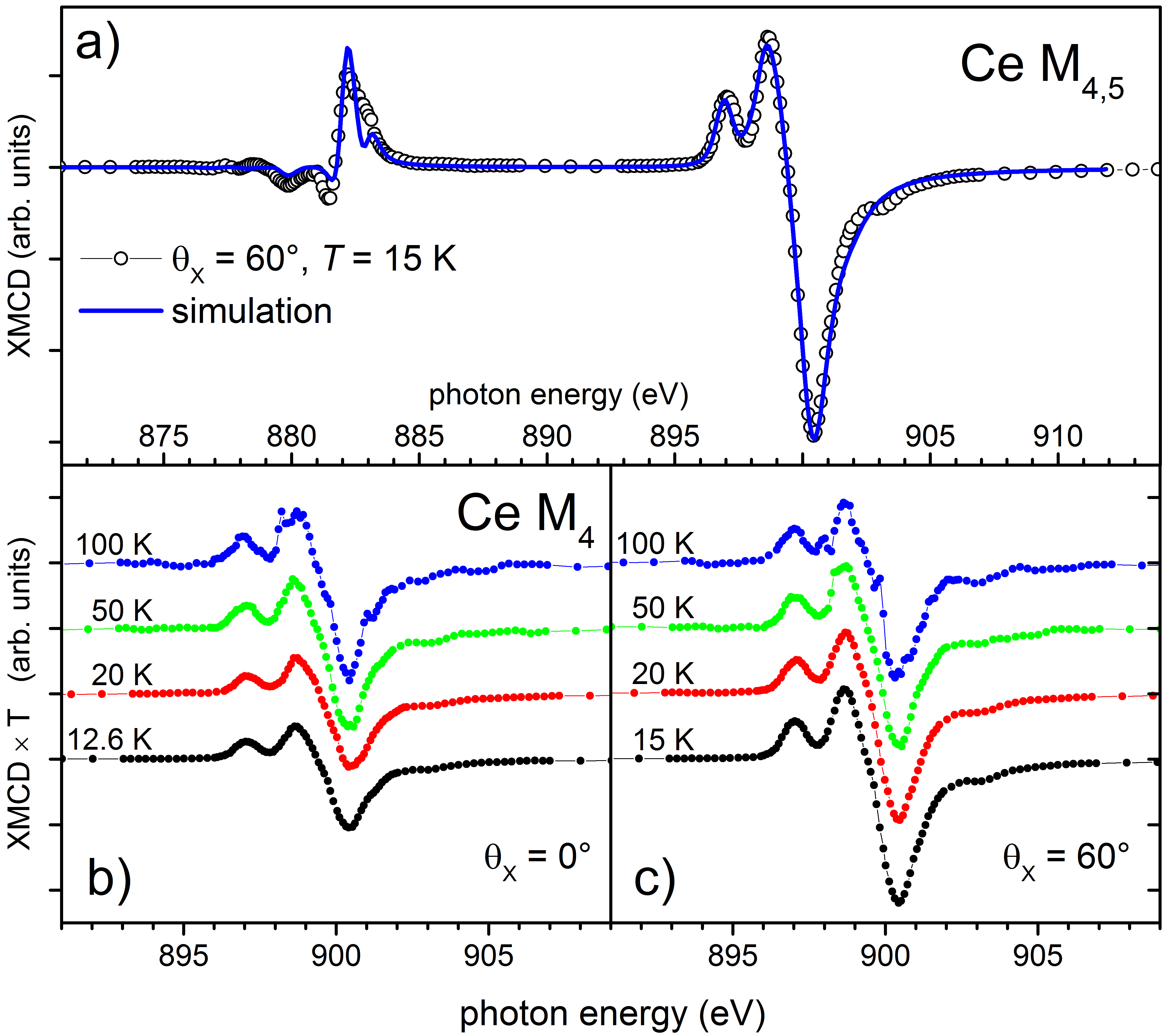} 
\caption{\label{fig3XMCD}
Selection of experimental and simulated XMCD spectra.
a) low temperature XMCD obtained at $\theta_X=60^\circ$ (``isotropic'' configuration) alongside with the calculated XMCD spectrum, scaled such as to match the magnitude of the experimental data.
b) temperature dependent Ce M$_4$ XMCD in NI geometry
c) same as in b) but for $\theta_X=60^\circ$.
}
\end{figure} 

The paramagnetic Ce $4f$ response was probed in an applied magnetic field of $\mu_0H=1.5$ T.
While sufficiently small to warrant linear response, it causes an XMCD signal which can reliably be measured over a considerable temperature range.
Fig.~\ref{fig3XMCD}a) displays the dichroic spectrum for the case of largest magnetic polarization obtained in the present work, alongside with the simulated
XMCD spectrum.
We notice that the spectral appearance of the XMCD is well accounted for by the atomic calculations, which were solely optimized to match the isotropic spectrum.
A notable exception is once again the spectral feature at $903$ eV, which produces a small but distinct contribution to magnetic dichroism in the experiment,
but is buried in the dichroism produced by the main M$_4$ peak in the calculation.
As in our previous work \cite{Prae15a}, we determine the Ce $4f$ polarization by applying the orbital moment XMCD sum rule \cite{Thol92a}
and assuming the atomic relation $m_S = -m_L/4$ between spin and orbital contributions to the total $4f$ magnetic moment to hold.
In the case of Fig.~\ref{fig3XMCD}a), the Ce $4f$ polarization amounts to approx.~$0.13$ $\mu_B$/atom.
The corresponding asymmetry in the XA spectra is largest at the M$_4$ edge and amounts to about $3.7$\% of the TEY signal.

The temperature dependence of the magnetic response at normal and oblique incidence, respectively,
is shown in Fig.~\ref{fig3XMCD}b) and c).
Each dichroic spectrum is multiplied by the value of the temperature at which it was obtained.
In this way, perfect Curie behavior would be reflected by a constant XMCD magnitude in the plots.
At low temperature in particular, the occurrence of single ion magnetic anisotropy is obvious.
Its sign and magnitude are directly related to the crystal field splitting scheme.
XMCD data therefore provide an independent probe of the CF scheme within the same set of experiments.


\begin{figure} 
\includegraphics[width=8.5cm]{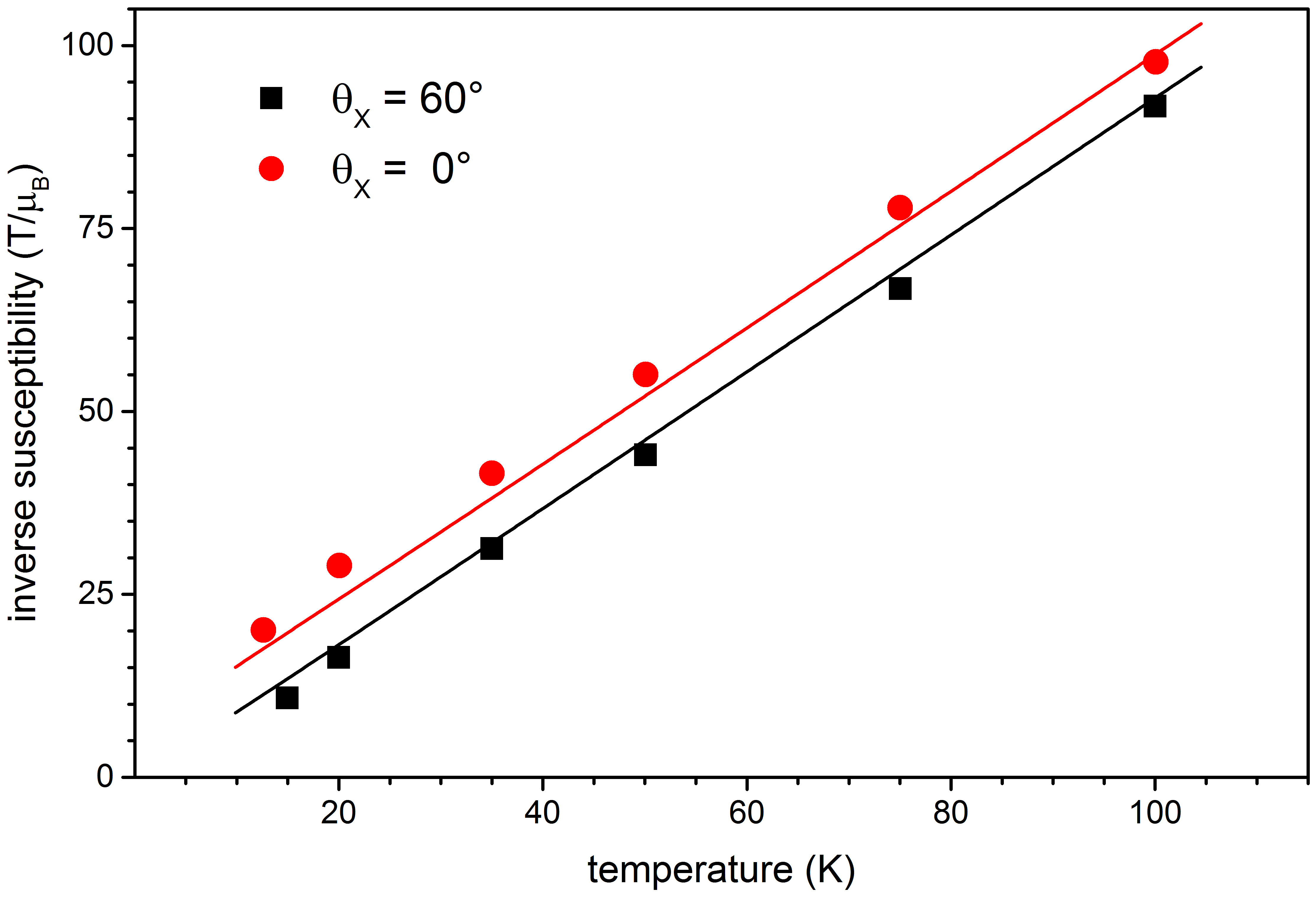} 
\caption{\label{fig4ichi}
Inverse magnetic Ce $4f$ susceptibilities of CeAg$_x$, determined from XMCD measurements, along with simulations
according to eq.~\ref{fullmodel}. CF parameters are the same as in Fig.~\ref{fig2XAS}d).
}
\end{figure} 

A more quantitative evaluation can be performed after extracting the temperature dependent Ce $4f$ susceptibilities from the XMCD data \cite{Prae15a}.
To second order they are given  \cite{Made68a,Luek79a} by the following expressions for the magnetic field applied along ($\chi_{||}$) and perpendicular ($\chi_\perp$) to the hexagonal axis, respectively:
\begin{eqnarray}
\chi^{}_{||} &= &\frac{g^2\mu_B^2}{4k_BTZ}  \left (  1 + 9p_1 + 25p_2 \right ) \\ 
\chi^{}_{\perp} &= & \frac{g^2\mu_B^2}{4k_BTZ} \cdot \left ( 9 + \frac{16k_BT}{\Delta_1} +\right. \nonumber\\
                    & &+\left ( \frac{10k_BT}{\Delta_2-\Delta_1}-\frac{16k_BT}{\Delta_1} \right ) p_1-\left. \frac{10k_BT}{\Delta_2-\Delta_1}  p_2 \right )  
\end{eqnarray}
At an intermediate angle $\theta$ the susceptibility reads
\begin{equation} \label{fullmodel}
\chi^{}_\theta = \frac{\cos^2\theta}{\chi_{||}^{-1}-\lambda}+\frac{\sin^2\theta}{\chi_{\perp}^{-1}-\lambda},
\end{equation}
where we have additionally allowed for magnetic coupling between Ce sites at the mean field level ($\lambda$).

In Fig.~\ref{fig4ichi} we show the temperature dependence of the inverse Ce $4f$ susceptibilities.
The crystal field splitting induced anisotropy leads to an offset between $\chi_{||}^{-1}$ and $\chi_\perp^{-1}$ which is nearly constant in the temperature range spanned by our experiments.
It is notably well perceptible up to high temperatures, where the precise determination of linear dichroism in our experiment is quite challenging.
The magnitude of this offset sensitively depends on the CF excitation energies $\Delta_1$ and $\Delta_2$.
Nevertheless, the magnetic response in this temperature range is not sufficient to pinpoint the numerical values of $\Delta_1$, $\Delta_2$ and $\lambda$.
In conjunction with the restrictions on $\Delta_2$ from above, however, we can determine the combination of parameters which produces the best simultaneous agreement
with both linear and circular dichroism.

The outcome of the parameter optimization is displayed in both Fig.~\ref{fig2XAS}d) and Fig.~\ref{fig4ichi} for the thermal evolution of the NI XAS and the
inverse susceptibilities, respectively.
We obtain $\Delta_1 = 0 \pm 0.4$ meV, $\Delta_2 = 1.25 \pm 0.05$ meV and  an insignificantly small mean field coupling $\lambda$.
These are the parameter values used for the simulations displayed in Figs.~\ref{fig2XAS}d) and \ref{fig4ichi}.

The correct slopes in $\chi^{-1}(T)$ in Fig.~\ref{fig4ichi} are only obtained by allowing for an overall reduction of the Ce magnetic moment by about 13\% compared to the value expected for free Ce$^{3+}$ ions, however.
From numerical simulations within the simplified NCA scheme proposed by Zwicknagl et al.~\cite{Zwic90a} we estimate that a Kondo temperature of  $T_K \gtrsim 20$ K would be required to produce this
reduction by Kondo screening.
Such a high value for $T_K$ appears quite unlikely, though, considering the small $4f$ hybridization found in the XA spectra and the photoemission results by Schwab et al.~\cite{Schw12a}. 
It is most likely, therefore, that the discrepancy is mostly due to an underestimation of the Ce $4f$ orbital moment in the sum rule evaluation.


In conclusion, we have presented a soft x-ray absorption study of an ultrathin, ordered intermetallic phase induced by alloying a sub-monolayer quantity of Ce into the surface of Ag(111).
Exploring the temperature dependences of both, linear and circular x-ray dichroism, we show that it is possible to explore the energy scales which characterize this material.
In addition to the smallness of $T_K$ \cite{Schw12a}, we find that both CF splittings and magnetic exchange coupling occur on energy scales of about 1 meV and below.
Our findings highlight both the enormous sensitivity of soft x-ray absorption and the usefulness of scheduling XMCD experiments when characterizing rare earth systems with soft x-rays. 
This extension comes at little cost, since many soft x-ray end stations provide the means to perform both linear and circular dichroism measurements.

\begin{acknowledgments}
We wish to thank H. Schwab and F. Reinert for helpful discussions as well as H.~Kie\ss ling and B.~M\"unzing for assistance with experiments.
This work was funded by the Deutsche Forschungsgemeinschaft through FOR1162.
Access to synchrotron radiation was partially granted by HZB. We also gratefully acknowledge HZB staff for their support during beam time. 
\end{acknowledgments}


\begin{thebibliography}{34}%
\makeatletter
\providecommand \@ifxundefined [1]{%
 \@ifx{#1\undefined}
}%
\providecommand \@ifnum [1]{%
 \ifnum #1\expandafter \@firstoftwo
 \else \expandafter \@secondoftwo
 \fi
}%
\providecommand \@ifx [1]{%
 \ifx #1\expandafter \@firstoftwo
 \else \expandafter \@secondoftwo
 \fi
}%
\providecommand \natexlab [1]{#1}%
\providecommand \enquote  [1]{``#1''}%
\providecommand \bibnamefont  [1]{#1}%
\providecommand \bibfnamefont [1]{#1}%
\providecommand \citenamefont [1]{#1}%
\providecommand \href@noop [0]{\@secondoftwo}%
\providecommand \href [0]{\begingroup \@sanitize@url \@href}%
\providecommand \@href[1]{\@@startlink{#1}\@@href}%
\providecommand \@@href[1]{\endgroup#1\@@endlink}%
\providecommand \@sanitize@url [0]{\catcode `\\12\catcode `\$12\catcode
  `\&12\catcode `\#12\catcode `\^12\catcode `\_12\catcode `\%12\relax}%
\providecommand \@@startlink[1]{}%
\providecommand \@@endlink[0]{}%
\providecommand \url  [0]{\begingroup\@sanitize@url \@url }%
\providecommand \@url [1]{\endgroup\@href {#1}{\urlprefix }}%
\providecommand \urlprefix  [0]{URL }%
\providecommand \Eprint [0]{\href }%
\providecommand \doibase [0]{http://dx.doi.org/}%
\providecommand \selectlanguage [0]{\@gobble}%
\providecommand \bibinfo  [0]{\@secondoftwo}%
\providecommand \bibfield  [0]{\@secondoftwo}%
\providecommand \translation [1]{[#1]}%
\providecommand \BibitemOpen [0]{}%
\providecommand \bibitemStop [0]{}%
\providecommand \bibitemNoStop [0]{.\EOS\space}%
\providecommand \EOS [0]{\spacefactor3000\relax}%
\providecommand \BibitemShut  [1]{\csname bibitem#1\endcsname}%
\let\auto@bib@innerbib\@empty
\bibitem [{\citenamefont {Gegenwart}\ \emph {et~al.}(2008)\citenamefont
  {Gegenwart}, \citenamefont {Si},\ and\ \citenamefont {Steglich}}]{Gege08a}%
  \BibitemOpen
  \bibfield  {author} {\bibinfo {author} {\bibfnamefont {P.}~\bibnamefont
  {Gegenwart}}, \bibinfo {author} {\bibfnamefont {Q.}~\bibnamefont {Si}}, \
  and\ \bibinfo {author} {\bibfnamefont {F.}~\bibnamefont {Steglich}},\
  }\href@noop {} {\bibfield  {journal} {\bibinfo  {journal} {Nat. Phys.}\
  }\textbf {\bibinfo {volume} {4}},\ \bibinfo {pages} {186} (\bibinfo {year}
  {2008})}\BibitemShut {NoStop}%
\bibitem [{\citenamefont {Gegenwart}\ \emph {et~al.}(2015)\citenamefont
  {Gegenwart}, \citenamefont {Steglich}, \citenamefont {Geibel},\ and\
  \citenamefont {Brando}}]{Gege15a}%
  \BibitemOpen
  \bibfield  {author} {\bibinfo {author} {\bibfnamefont {P.}~\bibnamefont
  {Gegenwart}}, \bibinfo {author} {\bibfnamefont {F.}~\bibnamefont {Steglich}},
  \bibinfo {author} {\bibfnamefont {C.}~\bibnamefont {Geibel}}, \ and\ \bibinfo
  {author} {\bibfnamefont {M.}~\bibnamefont {Brando}},\ }\href@noop {}
  {\bibfield  {journal} {\bibinfo  {journal} {Eur. Phys. J. Special Topics}\
  }\textbf {\bibinfo {volume} {224}},\ \bibinfo {pages} {975} (\bibinfo {year}
  {2015})}\BibitemShut {NoStop}%
\bibitem [{\citenamefont {Grewe}\ and\ \citenamefont
  {Steglich}(1991)}]{grew91a}%
  \BibitemOpen
  \bibfield  {author} {\bibinfo {author} {\bibfnamefont {N.}~\bibnamefont
  {Grewe}}\ and\ \bibinfo {author} {\bibfnamefont {F.}~\bibnamefont
  {Steglich}},\ }in\ \href@noop {} {\emph {\bibinfo {booktitle} {Handbook on
  the Physics and Chemistry of Rare Earths}}},\ Vol.~\bibinfo {volume} {14},\
  \bibinfo {editor} {edited by\ \bibinfo {editor} {\bibfnamefont {K.~A.}\
  \bibnamefont {{Gschneider, Jr.}}}\ and\ \bibinfo {editor} {\bibfnamefont
  {L.}~\bibnamefont {Eyring}}}\ (\bibinfo  {publisher} {Elsevier},\ \bibinfo
  {address} {Amsterdam},\ \bibinfo {year} {1991})\ p.\ \bibinfo {pages}
  {343}\BibitemShut {NoStop}%
\bibitem [{\citenamefont {von L\"ohneysen}\ \emph {et~al.}(2007)\citenamefont
  {von L\"ohneysen}, \citenamefont {Rosch}, \citenamefont {Vojta},\ and\
  \citenamefont {W\"olfle}}]{Lohn07a}%
  \BibitemOpen
  \bibfield  {author} {\bibinfo {author} {\bibfnamefont {H.}~\bibnamefont {von
  L\"ohneysen}}, \bibinfo {author} {\bibfnamefont {A.}~\bibnamefont {Rosch}},
  \bibinfo {author} {\bibfnamefont {M.}~\bibnamefont {Vojta}}, \ and\ \bibinfo
  {author} {\bibfnamefont {P.}~\bibnamefont {W\"olfle}},\ }\href {\doibase
  10.1103/RevModPhys.79.1015} {\bibfield  {journal} {\bibinfo  {journal} {Rev.
  Mod. Phys.}\ }\textbf {\bibinfo {volume} {79}},\ \bibinfo {pages} {1015}
  (\bibinfo {year} {2007})}\BibitemShut {NoStop}%
\bibitem [{\citenamefont {Yang}\ \emph {et~al.}(2008)\citenamefont {Yang},
  \citenamefont {Fisk}, \citenamefont {Lee}, \citenamefont {Thompson},\ and\
  \citenamefont {Pines}}]{yang08a}%
  \BibitemOpen
  \bibfield  {author} {\bibinfo {author} {\bibfnamefont {Y.}~\bibnamefont
  {Yang}}, \bibinfo {author} {\bibfnamefont {Z.}~\bibnamefont {Fisk}}, \bibinfo
  {author} {\bibfnamefont {H.-O.}\ \bibnamefont {Lee}}, \bibinfo {author}
  {\bibfnamefont {J.~D.}\ \bibnamefont {Thompson}}, \ and\ \bibinfo {author}
  {\bibfnamefont {D.}~\bibnamefont {Pines}},\ }\href@noop {} {\bibfield
  {journal} {\bibinfo  {journal} {Nature}\ }\textbf {\bibinfo {volume} {454}},\
  \bibinfo {pages} {611} (\bibinfo {year} {2008})}\BibitemShut {NoStop}%
\bibitem [{\citenamefont {Sacchi}\ \emph {et~al.}(1992)\citenamefont {Sacchi},
  \citenamefont {Sirotti},\ and\ \citenamefont {Rossi}}]{Sacc92a}%
  \BibitemOpen
  \bibfield  {author} {\bibinfo {author} {\bibfnamefont {M.}~\bibnamefont
  {Sacchi}}, \bibinfo {author} {\bibfnamefont {F.}~\bibnamefont {Sirotti}}, \
  and\ \bibinfo {author} {\bibfnamefont {G.}~\bibnamefont {Rossi}},\ }\href
  {\doibase 10.1016/0038-1098(92)90849-5} {\bibfield  {journal} {\bibinfo
  {journal} {Solid State Commun.}\ }\textbf {\bibinfo {volume} {81}},\ \bibinfo
  {pages} {977} (\bibinfo {year} {1992})}\BibitemShut {NoStop}%
\bibitem [{\citenamefont {Hansmann}\ \emph {et~al.}(2008)\citenamefont
  {Hansmann}, \citenamefont {Severing}, \citenamefont {Hu}, \citenamefont
  {Haverkort}, \citenamefont {Chang}, \citenamefont {Klein}, \citenamefont
  {Tanaka}, \citenamefont {Hsieh}, \citenamefont {Lin}, \citenamefont {Chen},
  \citenamefont {Fak}, \citenamefont {Lejay},\ and\ \citenamefont
  {Tjeng}}]{Hans08a}%
  \BibitemOpen
  \bibfield  {author} {\bibinfo {author} {\bibfnamefont {P.}~\bibnamefont
  {Hansmann}}, \bibinfo {author} {\bibfnamefont {A.}~\bibnamefont {Severing}},
  \bibinfo {author} {\bibfnamefont {Z.}~\bibnamefont {Hu}}, \bibinfo {author}
  {\bibfnamefont {M.~W.}\ \bibnamefont {Haverkort}}, \bibinfo {author}
  {\bibfnamefont {C.~F.}\ \bibnamefont {Chang}}, \bibinfo {author}
  {\bibfnamefont {S.}~\bibnamefont {Klein}}, \bibinfo {author} {\bibfnamefont
  {A.}~\bibnamefont {Tanaka}}, \bibinfo {author} {\bibfnamefont {H.~H.}\
  \bibnamefont {Hsieh}}, \bibinfo {author} {\bibfnamefont {H.~J.}\ \bibnamefont
  {Lin}}, \bibinfo {author} {\bibfnamefont {C.~T.}\ \bibnamefont {Chen}},
  \bibinfo {author} {\bibfnamefont {B.}~\bibnamefont {Fak}}, \bibinfo {author}
  {\bibfnamefont {P.}~\bibnamefont {Lejay}}, \ and\ \bibinfo {author}
  {\bibfnamefont {L.~H.}\ \bibnamefont {Tjeng}},\ }\href {\doibase
  10.1103/PhysRevLett.100.066405} {\bibfield  {journal} {\bibinfo  {journal}
  {Phys. Rev. Lett.}\ }\textbf {\bibinfo {volume} {100}},\ \bibinfo {pages}
  {066405} (\bibinfo {year} {2008})}\BibitemShut {NoStop}%
\bibitem [{\citenamefont {Willers}\ \emph {et~al.}(2009)\citenamefont
  {Willers}, \citenamefont {Fak}, \citenamefont {Hollmann}, \citenamefont
  {K{\"{o}}rner}, \citenamefont {Hu}, \citenamefont {Tanaka}, \citenamefont
  {Schmitz}, \citenamefont {Enderle}, \citenamefont {Lapertot}, \citenamefont
  {Tjeng},\ and\ \citenamefont {Severing}}]{Will09a}%
  \BibitemOpen
  \bibfield  {author} {\bibinfo {author} {\bibfnamefont {T.}~\bibnamefont
  {Willers}}, \bibinfo {author} {\bibfnamefont {B.}~\bibnamefont {Fak}},
  \bibinfo {author} {\bibfnamefont {N.}~\bibnamefont {Hollmann}}, \bibinfo
  {author} {\bibfnamefont {P.~O.}\ \bibnamefont {K{\"{o}}rner}}, \bibinfo {author}
  {\bibfnamefont {Z.}~\bibnamefont {Hu}}, \bibinfo {author} {\bibfnamefont
  {A.}~\bibnamefont {Tanaka}}, \bibinfo {author} {\bibfnamefont
  {D.}~\bibnamefont {Schmitz}}, \bibinfo {author} {\bibfnamefont
  {M.}~\bibnamefont {Enderle}}, \bibinfo {author} {\bibfnamefont
  {G.}~\bibnamefont {Lapertot}}, \bibinfo {author} {\bibfnamefont {L.~H.}\
  \bibnamefont {Tjeng}}, \ and\ \bibinfo {author} {\bibfnamefont
  {A.}~\bibnamefont {Severing}},\ }\href {\doibase 10.1103/PhysRevB.80.115106}
  {\bibfield  {journal} {\bibinfo  {journal} {Phys. Rev. B}\ }\textbf {\bibinfo
  {volume} {80}},\ \bibinfo {pages} {115106} (\bibinfo {year}
  {2009})}\BibitemShut {NoStop}%
\bibitem [{\citenamefont {Willers}\ \emph {et~al.}(2010)\citenamefont
  {Willers}, \citenamefont {Hu}, \citenamefont {Hollmann}, \citenamefont
  {K{\"{o}}rner}, \citenamefont {Gegner}, \citenamefont {Burnus}, \citenamefont
  {Fujiwara}, \citenamefont {Tanaka}, \citenamefont {Schmitz}, \citenamefont
  {Hsieh}, \citenamefont {Lin}, \citenamefont {Chen}, \citenamefont {Bauer},
  \citenamefont {Sarrao}, \citenamefont {Goremychkin}, \citenamefont {Koza},
  \citenamefont {Tjeng},\ and\ \citenamefont {Severing}}]{Will10a}%
  \BibitemOpen
  \bibfield  {author} {\bibinfo {author} {\bibfnamefont {T.}~\bibnamefont
  {Willers}}, \bibinfo {author} {\bibfnamefont {Z.}~\bibnamefont {Hu}},
  \bibinfo {author} {\bibfnamefont {N.}~\bibnamefont {Hollmann}}, \bibinfo
  {author} {\bibfnamefont {P.~O.}\ \bibnamefont {K{\"{o}}rner}}, \bibinfo {author}
  {\bibfnamefont {J.}~\bibnamefont {Gegner}}, \bibinfo {author} {\bibfnamefont
  {T.}~\bibnamefont {Burnus}}, \bibinfo {author} {\bibfnamefont
  {H.}~\bibnamefont {Fujiwara}}, \bibinfo {author} {\bibfnamefont
  {A.}~\bibnamefont {Tanaka}}, \bibinfo {author} {\bibfnamefont
  {D.}~\bibnamefont {Schmitz}}, \bibinfo {author} {\bibfnamefont {H.~H.}\
  \bibnamefont {Hsieh}}, \bibinfo {author} {\bibfnamefont {H.-J.}\ \bibnamefont
  {Lin}}, \bibinfo {author} {\bibfnamefont {C.~T.}\ \bibnamefont {Chen}},
  \bibinfo {author} {\bibfnamefont {E.~D.}\ \bibnamefont {Bauer}}, \bibinfo
  {author} {\bibfnamefont {J.~L.}\ \bibnamefont {Sarrao}}, \bibinfo {author}
  {\bibfnamefont {E.}~\bibnamefont {Goremychkin}}, \bibinfo {author}
  {\bibfnamefont {M.}~\bibnamefont {Koza}}, \bibinfo {author} {\bibfnamefont
  {L.~H.}\ \bibnamefont {Tjeng}}, \ and\ \bibinfo {author} {\bibfnamefont
  {A.}~\bibnamefont {Severing}},\ }\href {\doibase 10.1103/PhysRevB.81.195114}
  {\bibfield  {journal} {\bibinfo  {journal} {Phys. Rev. B}\ }\textbf {\bibinfo
  {volume} {81}},\ \bibinfo {pages} {195114} (\bibinfo {year}
  {2010})}\BibitemShut {NoStop}%
\bibitem [{\citenamefont {Willers}\ \emph {et~al.}(2011)\citenamefont
  {Willers}, \citenamefont {Cezar}, \citenamefont {Brookes}, \citenamefont
  {Hu}, \citenamefont {Strigari}, \citenamefont {K{\"{o}}rner}, \citenamefont
  {Hollmann}, \citenamefont {Schmitz}, \citenamefont {Bianchi}, \citenamefont
  {Fisk}, \citenamefont {Tanaka}, \citenamefont {Tjeng},\ and\ \citenamefont
  {Severing}}]{Will11a}%
  \BibitemOpen
  \bibfield  {author} {\bibinfo {author} {\bibfnamefont {T.}~\bibnamefont
  {Willers}}, \bibinfo {author} {\bibfnamefont {J.~C.}\ \bibnamefont {Cezar}},
  \bibinfo {author} {\bibfnamefont {N.~B.}\ \bibnamefont {Brookes}}, \bibinfo
  {author} {\bibfnamefont {Z.}~\bibnamefont {Hu}}, \bibinfo {author}
  {\bibfnamefont {F.}~\bibnamefont {Strigari}}, \bibinfo {author}
  {\bibfnamefont {P.}~\bibnamefont {K{\"{o}}rner}}, \bibinfo {author} {\bibfnamefont
  {N.}~\bibnamefont {Hollmann}}, \bibinfo {author} {\bibfnamefont
  {D.}~\bibnamefont {Schmitz}}, \bibinfo {author} {\bibfnamefont
  {A.}~\bibnamefont {Bianchi}}, \bibinfo {author} {\bibfnamefont
  {Z.}~\bibnamefont {Fisk}}, \bibinfo {author} {\bibfnamefont {A.}~\bibnamefont
  {Tanaka}}, \bibinfo {author} {\bibfnamefont {L.~H.}\ \bibnamefont {Tjeng}}, \
  and\ \bibinfo {author} {\bibfnamefont {A.}~\bibnamefont {Severing}},\ }\href
  {\doibase 10.1103/PhysRevLett.107.236402} {\bibfield  {journal} {\bibinfo
  {journal} {Phys. Rev. Lett.}\ }\textbf {\bibinfo {volume} {107}},\ \bibinfo
  {pages} {236402} (\bibinfo {year} {2011})}\BibitemShut {NoStop}%
\bibitem [{\citenamefont {Willers}\ \emph
  {et~al.}(2012{\natexlab{a}})\citenamefont {Willers}, \citenamefont {Adroja},
  \citenamefont {Rainford}, \citenamefont {Hu}, \citenamefont {Hollmann},
  \citenamefont {K{\"{o}}rner}, \citenamefont {Chin}, \citenamefont {Schmitz},
  \citenamefont {Hsieh}, \citenamefont {Lin}, \citenamefont {Chen},
  \citenamefont {Bauer}, \citenamefont {Sarrao}, \citenamefont {McClellan},
  \citenamefont {Byler}, \citenamefont {Geibel}, \citenamefont {Steglich},
  \citenamefont {Aoki}, \citenamefont {Lejay}, \citenamefont {Tanaka},
  \citenamefont {Tjeng},\ and\ \citenamefont {Severing}}]{Will12a}%
  \BibitemOpen
  \bibfield  {author} {\bibinfo {author} {\bibfnamefont {T.}~\bibnamefont
  {Willers}}, \bibinfo {author} {\bibfnamefont {D.~T.}\ \bibnamefont {Adroja}},
  \bibinfo {author} {\bibfnamefont {B.~D.}\ \bibnamefont {Rainford}}, \bibinfo
  {author} {\bibfnamefont {Z.}~\bibnamefont {Hu}}, \bibinfo {author}
  {\bibfnamefont {N.}~\bibnamefont {Hollmann}}, \bibinfo {author}
  {\bibfnamefont {P.~O.}\ \bibnamefont {K{\"{o}}rner}}, \bibinfo {author}
  {\bibfnamefont {Y.-Y.}\ \bibnamefont {Chin}}, \bibinfo {author}
  {\bibfnamefont {D.}~\bibnamefont {Schmitz}}, \bibinfo {author} {\bibfnamefont
  {H.~H.}\ \bibnamefont {Hsieh}}, \bibinfo {author} {\bibfnamefont {H.-J.}\
  \bibnamefont {Lin}}, \bibinfo {author} {\bibfnamefont {C.~T.}\ \bibnamefont
  {Chen}}, \bibinfo {author} {\bibfnamefont {E.~D.}\ \bibnamefont {Bauer}},
  \bibinfo {author} {\bibfnamefont {J.~L.}\ \bibnamefont {Sarrao}}, \bibinfo
  {author} {\bibfnamefont {K.~J.}\ \bibnamefont {McClellan}}, \bibinfo {author}
  {\bibfnamefont {D.}~\bibnamefont {Byler}}, \bibinfo {author} {\bibfnamefont
  {C.}~\bibnamefont {Geibel}}, \bibinfo {author} {\bibfnamefont
  {F.}~\bibnamefont {Steglich}}, \bibinfo {author} {\bibfnamefont
  {H.}~\bibnamefont {Aoki}}, \bibinfo {author} {\bibfnamefont {P.}~\bibnamefont
  {Lejay}}, \bibinfo {author} {\bibfnamefont {A.}~\bibnamefont {Tanaka}},
  \bibinfo {author} {\bibfnamefont {L.~H.}\ \bibnamefont {Tjeng}}, \ and\
  \bibinfo {author} {\bibfnamefont {A.}~\bibnamefont {Severing}},\ }\href
  {\doibase 10.1103/PhysRevB.85.035117} {\bibfield  {journal} {\bibinfo
  {journal} {Phys. Rev. B}\ }\textbf {\bibinfo {volume} {85}},\ \bibinfo
  {pages} {035117} (\bibinfo {year} {2012}{\natexlab{a}})}\BibitemShut
  {NoStop}%
\bibitem [{\citenamefont {Willers}\ \emph
  {et~al.}(2012{\natexlab{b}})\citenamefont {Willers}, \citenamefont
  {Strigari}, \citenamefont {Hiraoka}, \citenamefont {Cai}, \citenamefont
  {Haverkort}, \citenamefont {Tsuei}, \citenamefont {Liao}, \citenamefont
  {Seiro}, \citenamefont {Geibel}, \citenamefont {Steglich}, \citenamefont
  {Tjeng},\ and\ \citenamefont {Severing}}]{Will12b}%
  \BibitemOpen
  \bibfield  {author} {\bibinfo {author} {\bibfnamefont {T.}~\bibnamefont
  {Willers}}, \bibinfo {author} {\bibfnamefont {F.}~\bibnamefont {Strigari}},
  \bibinfo {author} {\bibfnamefont {N.}~\bibnamefont {Hiraoka}}, \bibinfo
  {author} {\bibfnamefont {Y.~Q.}\ \bibnamefont {Cai}}, \bibinfo {author}
  {\bibfnamefont {M.~W.}\ \bibnamefont {Haverkort}}, \bibinfo {author}
  {\bibfnamefont {K.-D.}\ \bibnamefont {Tsuei}}, \bibinfo {author}
  {\bibfnamefont {Y.~F.}\ \bibnamefont {Liao}}, \bibinfo {author}
  {\bibfnamefont {S.}~\bibnamefont {Seiro}}, \bibinfo {author} {\bibfnamefont
  {C.}~\bibnamefont {Geibel}}, \bibinfo {author} {\bibfnamefont
  {F.}~\bibnamefont {Steglich}}, \bibinfo {author} {\bibfnamefont {L.~H.}\
  \bibnamefont {Tjeng}}, \ and\ \bibinfo {author} {\bibfnamefont
  {A.}~\bibnamefont {Severing}},\ }\href {\doibase
  10.1103/PhysRevLett.109.046401} {\bibfield  {journal} {\bibinfo  {journal}
  {Phys. Rev. Lett.}\ }\textbf {\bibinfo {volume} {109}},\ \bibinfo {pages}
  {046401} (\bibinfo {year} {2012}{\natexlab{b}})}\BibitemShut {NoStop}%
\bibitem [{\citenamefont {Strigari}\ \emph {et~al.}(2012)\citenamefont
  {Strigari}, \citenamefont {Willers}, \citenamefont {Muro}, \citenamefont
  {Yutani}, \citenamefont {Takabatake}, \citenamefont {Hu}, \citenamefont
  {Chin}, \citenamefont {Agrestini}, \citenamefont {Lin}, \citenamefont {Chen},
  \citenamefont {Tanaka}, \citenamefont {Haverkort}, \citenamefont {Tjeng},\
  and\ \citenamefont {Severing}}]{Strig12a}%
  \BibitemOpen
  \bibfield  {author} {\bibinfo {author} {\bibfnamefont {F.}~\bibnamefont
  {Strigari}}, \bibinfo {author} {\bibfnamefont {T.}~\bibnamefont {Willers}},
  \bibinfo {author} {\bibfnamefont {Y.}~\bibnamefont {Muro}}, \bibinfo {author}
  {\bibfnamefont {K.}~\bibnamefont {Yutani}}, \bibinfo {author} {\bibfnamefont
  {T.}~\bibnamefont {Takabatake}}, \bibinfo {author} {\bibfnamefont
  {Z.}~\bibnamefont {Hu}}, \bibinfo {author} {\bibfnamefont {Y.-Y.}\
  \bibnamefont {Chin}}, \bibinfo {author} {\bibfnamefont {S.}~\bibnamefont
  {Agrestini}}, \bibinfo {author} {\bibfnamefont {H.-J.}\ \bibnamefont {Lin}},
  \bibinfo {author} {\bibfnamefont {C.~T.}\ \bibnamefont {Chen}}, \bibinfo
  {author} {\bibfnamefont {A.}~\bibnamefont {Tanaka}}, \bibinfo {author}
  {\bibfnamefont {M.~W.}\ \bibnamefont {Haverkort}}, \bibinfo {author}
  {\bibfnamefont {L.~H.}\ \bibnamefont {Tjeng}}, \ and\ \bibinfo {author}
  {\bibfnamefont {A.}~\bibnamefont {Severing}},\ }\href {\doibase
  10.1103/PhysRevB.86.081105} {\bibfield  {journal} {\bibinfo  {journal} {Phys.
  Rev. B}\ }\textbf {\bibinfo {volume} {86}},\ \bibinfo {pages} {081105}
  (\bibinfo {year} {2012})}\BibitemShut {NoStop}%
\bibitem [{\citenamefont {Willers}\ \emph {et~al.}(2015)\citenamefont
  {Willers}, \citenamefont {Strigari}, \citenamefont {Hu}, \citenamefont
  {Sessi}, \citenamefont {Brookes}, \citenamefont {Bauer}, \citenamefont
  {Sarrao}, \citenamefont {Thompson}, \citenamefont {Tanaka}, \citenamefont
  {Wirth}, \citenamefont {Tjeng},\ and\ \citenamefont {Severing}}]{Will15a}%
  \BibitemOpen
  \bibfield  {author} {\bibinfo {author} {\bibfnamefont {T.}~\bibnamefont
  {Willers}}, \bibinfo {author} {\bibfnamefont {F.}~\bibnamefont {Strigari}},
  \bibinfo {author} {\bibfnamefont {Z.}~\bibnamefont {Hu}}, \bibinfo {author}
  {\bibfnamefont {V.}~\bibnamefont {Sessi}}, \bibinfo {author} {\bibfnamefont
  {N.~B.}\ \bibnamefont {Brookes}}, \bibinfo {author} {\bibfnamefont {E.~D.}\
  \bibnamefont {Bauer}}, \bibinfo {author} {\bibfnamefont {J.~L.}\ \bibnamefont
  {Sarrao}}, \bibinfo {author} {\bibfnamefont {J.~D.}\ \bibnamefont
  {Thompson}}, \bibinfo {author} {\bibfnamefont {A.}~\bibnamefont {Tanaka}},
  \bibinfo {author} {\bibfnamefont {S.}~\bibnamefont {Wirth}}, \bibinfo
  {author} {\bibfnamefont {L.~H.}\ \bibnamefont {Tjeng}}, \ and\ \bibinfo
  {author} {\bibfnamefont {A.}~\bibnamefont {Severing}},\ }\href {\doibase
  10.1073/pnas.1415657112} {\bibfield  {journal} {\bibinfo  {journal}
  {Proceedings of the National Academy of Sciences}\ }\textbf {\bibinfo
  {volume} {112}},\ \bibinfo {pages} {2384} (\bibinfo {year} {2015})},\ \Eprint
  {http://arxiv.org/abs/http://www.pnas.org/content/112/8/2384.full.pdf}
  {http://www.pnas.org/content/112/8/2384.full.pdf} \BibitemShut {NoStop}%
\bibitem [{\citenamefont {Sch\"utz}\ \emph {et~al.}(1987)\citenamefont
  {Sch\"utz}, \citenamefont {Wagner}, \citenamefont {Wilhelm}, \citenamefont
  {Kienle}, \citenamefont {Zeller}, \citenamefont {Frahm},\ and\ \citenamefont
  {Materlik}}]{Schu87a}%
  \BibitemOpen
  \bibfield  {author} {\bibinfo {author} {\bibfnamefont {G.}~\bibnamefont
  {Sch\"utz}}, \bibinfo {author} {\bibfnamefont {W.}~\bibnamefont {Wagner}},
  \bibinfo {author} {\bibfnamefont {W.}~\bibnamefont {Wilhelm}}, \bibinfo
  {author} {\bibfnamefont {P.}~\bibnamefont {Kienle}}, \bibinfo {author}
  {\bibfnamefont {R.}~\bibnamefont {Zeller}}, \bibinfo {author} {\bibfnamefont
  {R.}~\bibnamefont {Frahm}}, \ and\ \bibinfo {author} {\bibfnamefont
  {G.}~\bibnamefont {Materlik}},\ }\href {\doibase 10.1103/PhysRevLett.58.737}
  {\bibfield  {journal} {\bibinfo  {journal} {Phys. Rev. Lett.}\ }\textbf
  {\bibinfo {volume} {58}},\ \bibinfo {pages} {737} (\bibinfo {year}
  {1987})}\BibitemShut {NoStop}%
\bibitem [{\citenamefont {van~der Laan}\ \emph {et~al.}(1986)\citenamefont
  {van~der Laan}, \citenamefont {Thole}, \citenamefont {Sawatzky},
  \citenamefont {Goedkoop}, \citenamefont {Fuggle}, \citenamefont {Esteva},
  \citenamefont {Karnatak}, \citenamefont {Remeika},\ and\ \citenamefont
  {Dabkowska}}]{Laan86b}%
  \BibitemOpen
  \bibfield  {author} {\bibinfo {author} {\bibfnamefont {G.}~\bibnamefont
  {van~der Laan}}, \bibinfo {author} {\bibfnamefont {B.~T.}\ \bibnamefont
  {Thole}}, \bibinfo {author} {\bibfnamefont {G.~A.}\ \bibnamefont {Sawatzky}},
  \bibinfo {author} {\bibfnamefont {J.~B.}\ \bibnamefont {Goedkoop}}, \bibinfo
  {author} {\bibfnamefont {J.~C.}\ \bibnamefont {Fuggle}}, \bibinfo {author}
  {\bibfnamefont {J.-M.}\ \bibnamefont {Esteva}}, \bibinfo {author}
  {\bibfnamefont {R.}~\bibnamefont {Karnatak}}, \bibinfo {author}
  {\bibfnamefont {J.~P.}\ \bibnamefont {Remeika}}, \ and\ \bibinfo {author}
  {\bibfnamefont {H.~A.}\ \bibnamefont {Dabkowska}},\ }\href {\doibase
  10.1103/PhysRevB.34.6529} {\bibfield  {journal} {\bibinfo  {journal} {Phys.
  Rev. B}\ }\textbf {\bibinfo {volume} {34}},\ \bibinfo {pages} {6529}
  (\bibinfo {year} {1986})}\BibitemShut {NoStop}%
\bibitem [{\citenamefont {Chen}\ \emph {et~al.}(1995)\citenamefont {Chen},
  \citenamefont {Idzerda}, \citenamefont {Lin}, \citenamefont {Smith},
  \citenamefont {Meigs}, \citenamefont {Chaban}, \citenamefont {Ho},
  \citenamefont {Pellegrin},\ and\ \citenamefont {Sette}}]{Chen95a}%
  \BibitemOpen
  \bibfield  {author} {\bibinfo {author} {\bibfnamefont {C.~T.}\ \bibnamefont
  {Chen}}, \bibinfo {author} {\bibfnamefont {Y.~U.}\ \bibnamefont {Idzerda}},
  \bibinfo {author} {\bibfnamefont {H.-J.}\ \bibnamefont {Lin}}, \bibinfo
  {author} {\bibfnamefont {N.~V.}\ \bibnamefont {Smith}}, \bibinfo {author}
  {\bibfnamefont {G.}~\bibnamefont {Meigs}}, \bibinfo {author} {\bibfnamefont
  {E.}~\bibnamefont {Chaban}}, \bibinfo {author} {\bibfnamefont {G.~H.}\
  \bibnamefont {Ho}}, \bibinfo {author} {\bibfnamefont {E.}~\bibnamefont
  {Pellegrin}}, \ and\ \bibinfo {author} {\bibfnamefont {F.}~\bibnamefont
  {Sette}},\ }\href {\doibase 10.1103/PhysRevLett.75.152} {\bibfield  {journal}
  {\bibinfo  {journal} {Phys. Rev. Lett.}\ }\textbf {\bibinfo {volume} {75}},\
  \bibinfo {pages} {152} (\bibinfo {year} {1995})}\BibitemShut {NoStop}%
\bibitem [{\citenamefont {St{\"{o}}hr}(1999)}]{stoe99a}%
  \BibitemOpen
  \bibfield  {author} {\bibinfo {author} {\bibfnamefont {J.}~\bibnamefont
  {St{\"{o}}hr}},\ }\href {\doibase
  http://dx.doi.org/10.1016/S0304-8853(99)00407-2} {\bibfield  {journal}
  {\bibinfo  {journal} {Journal of Magnetism and Magnetic Materials}\ }\textbf
  {\bibinfo {volume} {200}},\ \bibinfo {pages} {470 } (\bibinfo {year}
  {1999})}\BibitemShut {NoStop}%
\bibitem [{\citenamefont {van Elp}\ and\ \citenamefont
  {Searle}(1997)}]{vanE97a}%
  \BibitemOpen
  \bibfield  {author} {\bibinfo {author} {\bibfnamefont {J.}~\bibnamefont {van
  Elp}}\ and\ \bibinfo {author} {\bibfnamefont {B.}~\bibnamefont {Searle}},\
  }\href {\doibase http://dx.doi.org/10.1016/S0368-2048(97)00051-0} {\bibfield
  {journal} {\bibinfo  {journal} {J. Electron Spectrosc. Relat. Phenom.}\
  }\textbf {\bibinfo {volume} {86}},\ \bibinfo {pages} {93 } (\bibinfo {year}
  {1997})}\BibitemShut {NoStop}%
\bibitem [{\citenamefont {van~der Laan}\ \emph {et~al.}(2008)\citenamefont
  {van~der Laan}, \citenamefont {Arenholz}, \citenamefont {Schmehl},\ and\
  \citenamefont {Schlom}}]{vand08a}%
  \BibitemOpen
  \bibfield  {author} {\bibinfo {author} {\bibfnamefont {G.}~\bibnamefont
  {van~der Laan}}, \bibinfo {author} {\bibfnamefont {E.}~\bibnamefont
  {Arenholz}}, \bibinfo {author} {\bibfnamefont {A.}~\bibnamefont {Schmehl}}, \
  and\ \bibinfo {author} {\bibfnamefont {D.~G.}\ \bibnamefont {Schlom}},\
  }\href {\doibase 10.1103/PhysRevLett.100.067403} {\bibfield  {journal}
  {\bibinfo  {journal} {Phys. Rev. Lett.}\ }\textbf {\bibinfo {volume} {100}},\
  \bibinfo {pages} {067403} (\bibinfo {year} {2008})}\BibitemShut {NoStop}%
\bibitem [{\citenamefont {Praetorius}\ \emph {et~al.}(2015)\citenamefont
  {Praetorius}, \citenamefont {Zinner}, \citenamefont {K\"ohl}, \citenamefont
  {Kie\ss{}ling}, \citenamefont {Br\"uck}, \citenamefont {Muenzing},
  \citenamefont {Kamp}, \citenamefont {Kachel}, \citenamefont {Choueikani},
  \citenamefont {Ohresser}, \citenamefont {Wilhelm}, \citenamefont {Rogalev},\
  and\ \citenamefont {Fauth}}]{Prae15a}%
  \BibitemOpen
  \bibfield  {author} {\bibinfo {author} {\bibfnamefont {C.}~\bibnamefont
  {Praetorius}}, \bibinfo {author} {\bibfnamefont {M.}~\bibnamefont {Zinner}},
  \bibinfo {author} {\bibfnamefont {A.}~\bibnamefont {K\"ohl}}, \bibinfo
  {author} {\bibfnamefont {H.}~\bibnamefont {Kie\ss{}ling}}, \bibinfo {author}
  {\bibfnamefont {S.}~\bibnamefont {Br\"uck}}, \bibinfo {author} {\bibfnamefont
  {B.}~\bibnamefont {Muenzing}}, \bibinfo {author} {\bibfnamefont
  {M.}~\bibnamefont {Kamp}}, \bibinfo {author} {\bibfnamefont {T.}~\bibnamefont
  {Kachel}}, \bibinfo {author} {\bibfnamefont {F.}~\bibnamefont {Choueikani}},
  \bibinfo {author} {\bibfnamefont {P.}~\bibnamefont {Ohresser}}, \bibinfo
  {author} {\bibfnamefont {F.}~\bibnamefont {Wilhelm}}, \bibinfo {author}
  {\bibfnamefont {A.}~\bibnamefont {Rogalev}}, \ and\ \bibinfo {author}
  {\bibfnamefont {K.}~\bibnamefont {Fauth}},\ }\href {\doibase
  10.1103/PhysRevB.92.045116} {\bibfield  {journal} {\bibinfo  {journal} {Phys.
  Rev. B}\ }\textbf {\bibinfo {volume} {92}},\ \bibinfo {pages} {045116}
  (\bibinfo {year} {2015})}\BibitemShut {NoStop}%
\bibitem [{\citenamefont {Schwab}\ \emph {et~al.}(2012)\citenamefont {Schwab},
  \citenamefont {Mulazzi}, \citenamefont {Jiang}, \citenamefont {Hayashi},
  \citenamefont {Habuchi}, \citenamefont {Hirayama}, \citenamefont {Iwasawa},
  \citenamefont {Shimada},\ and\ \citenamefont {Reinert}}]{Schw12a}%
  \BibitemOpen
  \bibfield  {author} {\bibinfo {author} {\bibfnamefont {H.}~\bibnamefont
  {Schwab}}, \bibinfo {author} {\bibfnamefont {M.}~\bibnamefont {Mulazzi}},
  \bibinfo {author} {\bibfnamefont {J.}~\bibnamefont {Jiang}}, \bibinfo
  {author} {\bibfnamefont {H.}~\bibnamefont {Hayashi}}, \bibinfo {author}
  {\bibfnamefont {T.}~\bibnamefont {Habuchi}}, \bibinfo {author} {\bibfnamefont
  {D.}~\bibnamefont {Hirayama}}, \bibinfo {author} {\bibfnamefont
  {H.}~\bibnamefont {Iwasawa}}, \bibinfo {author} {\bibfnamefont
  {K.}~\bibnamefont {Shimada}}, \ and\ \bibinfo {author} {\bibfnamefont
  {F.}~\bibnamefont {Reinert}},\ }\href {\doibase 10.1103/PhysRevB.85.125130}
  {\bibfield  {journal} {\bibinfo  {journal} {Phys. Rev. B}\ }\textbf {\bibinfo
  {volume} {85}},\ \bibinfo {pages} {125130} (\bibinfo {year}
  {2012})}\BibitemShut {NoStop}%
\bibitem [{\citenamefont {Kemmer}\ \emph {et~al.}(2014)\citenamefont {Kemmer},
  \citenamefont {Praetorius}, \citenamefont {Kr\"onlein}, \citenamefont {Hsu},
  \citenamefont {Fauth},\ and\ \citenamefont {Bode}}]{Kemm14a}%
  \BibitemOpen
  \bibfield  {author} {\bibinfo {author} {\bibfnamefont {J.}~\bibnamefont
  {Kemmer}}, \bibinfo {author} {\bibfnamefont {C.}~\bibnamefont {Praetorius}},
  \bibinfo {author} {\bibfnamefont {A.}~\bibnamefont {Kr\"onlein}}, \bibinfo
  {author} {\bibfnamefont {P.-J.}\ \bibnamefont {Hsu}}, \bibinfo {author}
  {\bibfnamefont {K.}~\bibnamefont {Fauth}}, \ and\ \bibinfo {author}
  {\bibfnamefont {M.}~\bibnamefont {Bode}},\ }\href {\doibase
  10.1103/PhysRevB.90.195401} {\bibfield  {journal} {\bibinfo  {journal} {Phys.
  Rev. B}\ }\textbf {\bibinfo {volume} {90}},\ \bibinfo {pages} {195401}
  (\bibinfo {year} {2014})}\BibitemShut {NoStop}%
\bibitem [{\citenamefont {{Zinner et al.}}()}]{Zinnerunpub}%
  \BibitemOpen
  \bibfield  {author} {\bibinfo {author} {\bibfnamefont {M.}~\bibnamefont
  {{Zinner et al.}}},\ }\href@noop {} {}\bibinfo {note} {Unpublished
  results}\BibitemShut {NoStop}%
\bibitem [{\citenamefont {Kachel}\ \emph {et~al.}(2015)\citenamefont {Kachel},
  \citenamefont {Eggenstein},\ and\ \citenamefont {Follath}}]{Kach15a}%
  \BibitemOpen
  \bibfield  {author} {\bibinfo {author} {\bibfnamefont {T.}~\bibnamefont
  {Kachel}}, \bibinfo {author} {\bibfnamefont {F.}~\bibnamefont {Eggenstein}},
  \ and\ \bibinfo {author} {\bibfnamefont {R.}~\bibnamefont {Follath}},\ }\href
  {\doibase 10.1107/S1600577515010826} {\bibfield  {journal} {\bibinfo
  {journal} {Journal of Synchrotron Radiation}\ }\textbf {\bibinfo {volume}
  {22}},\ \bibinfo {pages} {1301} (\bibinfo {year} {2015})}\BibitemShut
  {NoStop}%
\bibitem [{\citenamefont {Haverkort}\ \emph {et~al.}(2012)\citenamefont
  {Haverkort}, \citenamefont {Zwierzycki},\ and\ \citenamefont
  {Andersen}}]{Have12a}%
  \BibitemOpen
  \bibfield  {author} {\bibinfo {author} {\bibfnamefont {M.~W.}\ \bibnamefont
  {Haverkort}}, \bibinfo {author} {\bibfnamefont {M.}~\bibnamefont
  {Zwierzycki}}, \ and\ \bibinfo {author} {\bibfnamefont {O.~K.}\ \bibnamefont
  {Andersen}},\ }\href@noop {} {\bibfield  {journal} {\bibinfo  {journal}
  {Phys. Rev. B}\ }\textbf {\bibinfo {volume} {85}},\ \bibinfo {pages} {165113}
  (\bibinfo {year} {2012})}\BibitemShut {NoStop}%
\bibitem [{\citenamefont {{Haverkort et al.}}()}]{quantylink}%
  \BibitemOpen
  \bibfield  {author} {\bibinfo {author} {\bibfnamefont {M.~W.}\ \bibnamefont
  {{Haverkort et al.}}},\ }\href {http://www.quanty.org} {\ }\bibinfo {note}
  {{http://www.quanty.org}}\BibitemShut {NoStop}%
\bibitem [{\citenamefont {Strigari}\ \emph {et~al.}(2013)\citenamefont
  {Strigari}, \citenamefont {Willers}, \citenamefont {Muro}, \citenamefont
  {Yutani}, \citenamefont {Takabatake}, \citenamefont {Hu}, \citenamefont
  {Agrestini}, \citenamefont {Kuo}, \citenamefont {Chin}, \citenamefont {Lin},
  \citenamefont {Pi}, \citenamefont {Chen}, \citenamefont {Weschke},
  \citenamefont {Schierle}, \citenamefont {Tanaka}, \citenamefont {Haverkort},
  \citenamefont {Tjeng},\ and\ \citenamefont {Severing}}]{Stri13a}%
  \BibitemOpen
  \bibfield  {author} {\bibinfo {author} {\bibfnamefont {F.}~\bibnamefont
  {Strigari}}, \bibinfo {author} {\bibfnamefont {T.}~\bibnamefont {Willers}},
  \bibinfo {author} {\bibfnamefont {Y.}~\bibnamefont {Muro}}, \bibinfo {author}
  {\bibfnamefont {K.}~\bibnamefont {Yutani}}, \bibinfo {author} {\bibfnamefont
  {T.}~\bibnamefont {Takabatake}}, \bibinfo {author} {\bibfnamefont
  {Z.}~\bibnamefont {Hu}}, \bibinfo {author} {\bibfnamefont {S.}~\bibnamefont
  {Agrestini}}, \bibinfo {author} {\bibfnamefont {C.-Y.}\ \bibnamefont {Kuo}},
  \bibinfo {author} {\bibfnamefont {Y.-Y.}\ \bibnamefont {Chin}}, \bibinfo
  {author} {\bibfnamefont {H.-J.}\ \bibnamefont {Lin}}, \bibinfo {author}
  {\bibfnamefont {T.~W.}\ \bibnamefont {Pi}}, \bibinfo {author} {\bibfnamefont
  {C.~T.}\ \bibnamefont {Chen}}, \bibinfo {author} {\bibfnamefont
  {E.}~\bibnamefont {Weschke}}, \bibinfo {author} {\bibfnamefont
  {E.}~\bibnamefont {Schierle}}, \bibinfo {author} {\bibfnamefont
  {A.}~\bibnamefont {Tanaka}}, \bibinfo {author} {\bibfnamefont {M.~W.}\
  \bibnamefont {Haverkort}}, \bibinfo {author} {\bibfnamefont {L.~H.}\
  \bibnamefont {Tjeng}}, \ and\ \bibinfo {author} {\bibfnamefont
  {A.}~\bibnamefont {Severing}},\ }\href {\doibase 10.1103/PhysRevB.87.125119}
  {\bibfield  {journal} {\bibinfo  {journal} {Phys. Rev. B}\ }\textbf {\bibinfo
  {volume} {87}},\ \bibinfo {pages} {125119} (\bibinfo {year}
  {2013})}\BibitemShut {NoStop}%
\bibitem [{\citenamefont {Thole}\ \emph {et~al.}(1985)\citenamefont {Thole},
  \citenamefont {van~der Laan}, \citenamefont {Fuggle}, \citenamefont
  {Sawatzky}, \citenamefont {Karnatak},\ and\ \citenamefont
  {Esteva}}]{Thol85a}%
  \BibitemOpen
  \bibfield  {author} {\bibinfo {author} {\bibfnamefont {B.~T.}\ \bibnamefont
  {Thole}}, \bibinfo {author} {\bibfnamefont {G.}~\bibnamefont {van~der Laan}},
  \bibinfo {author} {\bibfnamefont {J.~C.}\ \bibnamefont {Fuggle}}, \bibinfo
  {author} {\bibfnamefont {G.~A.}\ \bibnamefont {Sawatzky}}, \bibinfo {author}
  {\bibfnamefont {R.~C.}\ \bibnamefont {Karnatak}}, \ and\ \bibinfo {author}
  {\bibfnamefont {J.~M.}\ \bibnamefont {Esteva}},\ }\href {\doibase
  10.1103/PhysRevB.32.5107} {\bibfield  {journal} {\bibinfo  {journal} {Phys.
  Rev. B}\ }\textbf {\bibinfo {volume} {32}},\ \bibinfo {pages} {5107}
  (\bibinfo {year} {1985})}\BibitemShut {NoStop}%
\bibitem [{\citenamefont {Schippers}(2011)}]{Schi11a}%
  \BibitemOpen
  \bibfield  {author} {\bibinfo {author} {\bibfnamefont {S.}~\bibnamefont
  {Schippers}},\ }\href@noop {} {\bibfield  {journal} {\bibinfo  {journal}
  {Int. Rev. At. Mol. Phys.}\ }\textbf {\bibinfo {volume} {2}},\ \bibinfo {pages} {151} (\bibinfo
  {year} {2011})}\BibitemShut {NoStop}%
\bibitem [{\citenamefont {Thole}\ \emph {et~al.}(1992)\citenamefont {Thole},
  \citenamefont {Carra}, \citenamefont {Sette},\ and\ \citenamefont {van~der
  Laan}}]{Thol92a}%
  \BibitemOpen
  \bibfield  {author} {\bibinfo {author} {\bibfnamefont {B.~T.}\ \bibnamefont
  {Thole}}, \bibinfo {author} {\bibfnamefont {P.}~\bibnamefont {Carra}},
  \bibinfo {author} {\bibfnamefont {F.}~\bibnamefont {Sette}}, \ and\ \bibinfo
  {author} {\bibfnamefont {G.}~\bibnamefont {van~der Laan}},\ }\href {\doibase
  10.1103/PhysRevLett.68.1943} {\bibfield  {journal} {\bibinfo  {journal}
  {Phys. Rev. Lett.}\ }\textbf {\bibinfo {volume} {68}},\ \bibinfo {pages}
  {1943} (\bibinfo {year} {1992})}\BibitemShut {NoStop}%
\bibitem [{\citenamefont {Mader}\ and\ \citenamefont {Swift}(1968)}]{Made68a}%
  \BibitemOpen
  \bibfield  {author} {\bibinfo {author} {\bibfnamefont {K.~H.}\ \bibnamefont
  {Mader}}\ and\ \bibinfo {author} {\bibfnamefont {M.~W.}\ \bibnamefont
  {Swift}},\ }\href@noop {} {\bibfield  {journal} {\bibinfo  {journal} {J.
  Phys. Chem. Solids}\ }\textbf {\bibinfo {volume} {29}},\ \bibinfo {pages}
  {1759} (\bibinfo {year} {1968})}\BibitemShut {NoStop}%
\bibitem [{\citenamefont {Lueken}\ \emph {et~al.}(1979)\citenamefont {Lueken},
  \citenamefont {Meier}, \citenamefont {Klessen}, \citenamefont {Bronger},\
  and\ \citenamefont {Fleischhauer}}]{Luek79a}%
  \BibitemOpen
  \bibfield  {author} {\bibinfo {author} {\bibfnamefont {H.}~\bibnamefont
  {Lueken}}, \bibinfo {author} {\bibfnamefont {M.}~\bibnamefont {Meier}},
  \bibinfo {author} {\bibfnamefont {G.}~\bibnamefont {Klessen}}, \bibinfo
  {author} {\bibfnamefont {W.}~\bibnamefont {Bronger}}, \ and\ \bibinfo
  {author} {\bibfnamefont {J.}~\bibnamefont {Fleischhauer}},\ }\href {\doibase
  10.1016/0022-5088(79)90219-4} {\bibfield  {journal} {\bibinfo  {journal} {J.
  Less-Comm. Met.}\ }\textbf {\bibinfo {volume} {63}},\ \bibinfo {pages} {P35 }
  (\bibinfo {year} {1979})}\BibitemShut {NoStop}%
\bibitem [{\citenamefont {Zwicknagl}\ \emph {et~al.}(1990)\citenamefont
  {Zwicknagl}, \citenamefont {Zevin},\ and\ \citenamefont {Fulde}}]{Zwic90a}%
  \BibitemOpen
  \bibfield  {author} {\bibinfo {author} {\bibfnamefont {G.}~\bibnamefont
  {Zwicknagl}}, \bibinfo {author} {\bibfnamefont {V.}~\bibnamefont {Zevin}}, \
  and\ \bibinfo {author} {\bibfnamefont {P.}~\bibnamefont {Fulde}},\ }\href
  {\doibase 10.1007/BF01437646} {\bibfield  {journal} {\bibinfo  {journal} {Z.
  Phys. B}\ }\textbf {\bibinfo {volume} {79}},\ \bibinfo {pages} {365}
  (\bibinfo {year} {1990})}\BibitemShut {NoStop}%
\end{thebibliography}

%

\end{document}